\documentclass[twocolumn,showpacs,preprintnumbers,amsmath,amssymb,epsf]{revtex4-1}

\usepackage{graphicx,epsfig}
\usepackage{dcolumn}
\usepackage{bm}
\newcommand{\be}{\begin{equation}}
\newcommand{\ee}{\end{equation}}
\newcommand{\bse}{\begin{subequations}}
\newcommand{\ese}{\end{subequations}}
\newcommand{\bary}{\begin{eqnarray}}
\newcommand{\eary}{\end{eqnarray}}
\newcommand{\bwt}{\begin{widetext}}
\newcommand{\ewt}{\end{widetext}}

\begin{document}


\title{Hadronic-Origin TeV $\gamma$-Rays and Ultra-High Energy Cosmic Rays from
Centaurus A}
\author{Sarira Sahu$^{*}$, Bing Zhang$^{**}$ and Nissim Fraija$^{\dagger}$}
\affiliation{
$^{*}$Instituto de Ciencias Nucleares, Universidad Nacional Aut\'onoma de M\'exico, 
Circuito Exterior, C.U., A. Postal 70-543, 04510 Mexico DF, Mexico\\
$^{**}$Department of Physics and Astronomy, University of Nevada, Las Vegas, NV 89154, USA\\
$^{\dagger}$Instituto de Astronomia, Universidad Nacional Aut\'onoma de M\'exico, 
Circuito Exterior, C.U., A. Postal 70-543, 04510 Mexico DF, Mexico
}


\begin{abstract}

Centaurus A (Cen A) is the nearest radio-loud AGN and is detected from 
radio to very high energy gamma-rays.  Its nuclear spectral energy
distribution (SED) shows a double-peak feature, which is well explained
by the leptonic synchrotron + synchrotron self-Compton model. This model
however cannot account for the observed high energy photons in the TeV 
range, which display a distinct component.
Here we show that $\sim$ TeV photons can be well interpreted as
the $\pi^0$ decay products from $p\gamma$ interactions of 
Fermi accelerated high energy protons in the jet with the seed 
photons around the second SED peak at $\sim 170$ keV. 
Extrapolating the inferred proton spectrum to high energies, we find
that this same model 
 is consistent with the detection of 
2 ultra-high-energy cosmic ray events detected by Pierre Auger Observatory 
from the direction of
Cen A. 
We also estimate the GeV neutrino flux from the same process,
and find that it is too faint to be detected by current
high-energy neutrino detectors.

\end{abstract}

\maketitle

\begin{figure}[t!]
\vspace{0.3cm}
{\centering
\resizebox*{0.5\textwidth}{0.3\textheight}
{\includegraphics{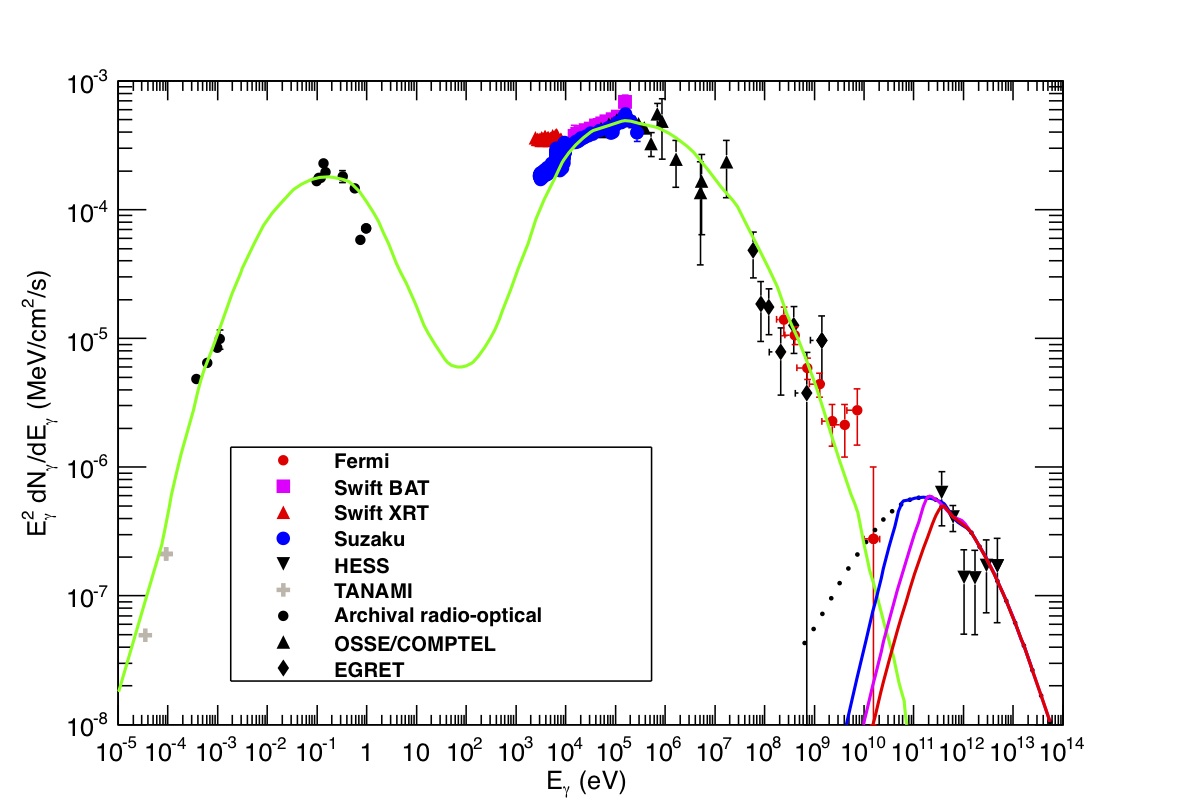}}
\par}
\caption{The observed spectral energy distribution
    $E^2_{\gamma}\frac{dN_{\gamma}}{dE_{\gamma}}$ (or $F_\gamma$) from the 
    Cen A core region. Colored symbol points are observations (with 
    different sources marked in the figure), and the curves are model fits:
    The green curve is synchrotron + SSC fit from Abdo et. al. [6],  
while 
    the blue, magenta, and red curves are the hadronic emission component 
    due to $\pi^0$-decay with $E_p^b=4, 13, 25$ TeV, respectively. The case
    of no proton spectral break (black dashed line) is plotted for
    comparison.
   }
\label{SED}
\end{figure}

{\em I. Introduction. ---}
Centaurus A (Cen A or NGC 5128) is the nearest active  radio galaxy
with a distance of approximately 3.5 Mpc and redshift 
$z=0.002$ \cite{Israel:1998ws}.  Although by active galactic nuclei (AGN) 
standard its bolometric 
luminosity is not very large, because of its proximity to earth it is one 
of the best studied AGN. Optically, Cen A is an elliptical galaxy
undergoing late stages of a merger event with a small 
spiral galaxy.  Sufficiently large amount of photometric
data are available to build a well sampled spectral energy 
distribution (SED) of Cen A.  The emission from the nucleus of Cen A has been
observed throughout the electromagnetic spectrum from radio to
gamma-rays \cite{Winkler,Hardcastle:2003ye,Sreekumar:1999xw,Aharonian:2009xn,Abdo:2010fk},
which shows that Cen A has a FR I morphology 
with two radio lobes, and is a non-blazar source with a jet inclination in
the range of $15^{\rm o}$ to $80^{\rm o}$.  The nuclear SED shows two peaks, one
in the far-infrared band ($\sim 4\times 10^{-2}$ eV) and another at around
170 keV \cite{Jourdain:1993}.  
In the framework of the unification scheme of AGN, blazars and radio 
galaxies  \cite{Ghisellini:1998it, Fossati:1998zn, Chiaberge:2001ek} are
intrinsically the same objects, viewed at different angles with respect
to the jet axis. The double-peak SED structure observed in Cen A
is similar to that of blazars whose jets beam towards earth, suggesting 
that the same spectral feature is also expected from misaligned jet 
sources such as Cen A. 
The leading interpretation is the single-zone synchrotron and
synchrotron self-Compton (SSC) model. In this scenario, the multi
wavelength emission originates from the same region. 
The low-energy emission in radio to optical wavelengths is the non-thermal
synchrotron radiation from a population of relativistic electrons in the
jet, while high-energy emission from X-rays to very high energy
(VHE) gamma-rays are from the Compton scattering of the above seed synchrotron
photons by the same population of electrons.
This model is found very successful
in explaining the multi-wavelength emission from BL Lac objects and FR I
galaxies such as NGC 1275 and M87  \cite{Ghisellini:1998it,Fossati:1998zn}.
Applying it to Cen A, one found that it is successful in explaining 
most of the multi wavelength SED data \cite{Abdo:2010fk,Roustazadeh:2011zz}.
The difficulty, on the other hand, is the multi-TeV emission detected by HESS
during 2004 to 2008 \cite{Aharonian:2009xn}.
Even though the HESS data alone can be fitted by a power law \cite{Aharonian:2009xn},
and the entire $10^6 - 10^{13}$ eV spectrum may be still accommodated within 
one single power law model, a clear dip around $10^{10}$ eV revealed by Fermi
indicates an excess of TeV emission from
the extrapolation of the Fermi data \cite{Abdo:2010fk} (Fig.\ref{SED}).
So far this TeV spectral component is not well interpreted within the published
leptonic models \cite{Abdo:2010fk,Roustazadeh:2011zz}.

On the other hand, Cen A has long been proposed as the source of very high
energy cosmic rays. Recently Pierre Auger Collaboration reported 
two UHECR events fall within $3.1^{\rm o}$ around Cen A \cite{Auger:2009}.
By assuming that the two events are from Cen A,
the expected high energy neutrino event rates in detectors such as
IceCube \cite{Halzen:2008vz,:2007qd} and the diffuse neutrino
flux from Cen A \cite{Koers:2008hv} have been estimated. The flux of
high energy cosmic rays as well as the accompanying expected secondary 
photons and neutrinos are calculated from hadronic models
\cite{Kachelriess:2008qx}. 

The astrophysical objects producing UHECRs also produce high 
energy $\gamma$-rays due to interaction of the cosmic rays with the 
background through $pp$ or $p\gamma$ interactions 
\cite{Kachelriess:2008qx,Romero:1995tn,Isola:2001ng,Honda:2009xd,Kachelriess:2009wn}.
In this letter, we show that the multi-TeV $\gamma$-ray emission from Cen A
can be naturally interpreted by a hadronic model invoking $p\gamma$
interactions between Fermi-accelerated protons in the jet and the seed photons 
near the SSC peak (170 keV). 
The same model is found consistent with the detection of 2 UHECR events 
from Cen A.

{\em II. Hadronic model of TeV gamma-rays.}
Observations of variable, non-thermal high energy emission from AGNs
imply that these sources are efficient accelerators of
particles through shock or diffusive Fermi acceleration processes.
While efficient electron acceleration is
limited by high radiative losses, protons and heavy nuclei can
reach UHE through the same acceleration mechanism. In
general, these  energetic charged particles (electrons and protons)
have a power-law spectrum given as ${dN}/{dE} \propto E^{-\alpha}$,
with the power index $\alpha \geq 2$ \cite{Dermer:1993cz, Gupta:2008tm}.

The dominant $p\gamma$ interaction is through the $\Delta$-resonance, i.e.
\be
p+\gamma \rightarrow \Delta^+\rightarrow  
 \left\{ 
\begin{array}{l l}
 p\,\pi^0, & \quad \text {fraction~ 2/3}\\
  n\,\pi^+  \rightarrow n e^{+}\nu_e\nu_\mu \bar\nu_\mu, 
& \quad  \text {fraction~ 1/3}\\
\end{array} \right. ,
\label{decaymode}
\ee
which has a cross section $\sigma_{\Delta}\sim 5\times 10^{-28}\,
cm^2$. The charged $\pi$'s subsequently decay to charged leptons and
neutrinos, while neutral $\pi$'s decay to GeV-TeV photons.
For interactions at $\Delta$-resonance, the matching condition is
$E'_p \epsilon'_\gamma \simeq 0.32
\,(1-\cos\theta)^{-1}\, {\rm GeV}^2$, 
where $E'_p$ and $\epsilon'_\gamma$ are
the proton and the background IC photon energies in the comoving frame
of the jet, respectively. 
Since in the comoving frame the protons collide with the IC photons
from all directions, 
in our calculation we consider an average value $(1-\cos\theta) \sim 1$ 
($\theta$ in the range of 0 and $\pi$). 
Denoting $\Gamma$ as the bulk Lorentz factor of the jet, and
${\cal D}=\Gamma^{-1} (1-\beta \cos\theta_{\rm ob})^{-1}$ as the 
Doppler factor to the observer ($\theta_{\rm ob}$
is the angle between the observer and the jet direction, and
$\beta = v/c$ is the dimensionless speed of the jet), 
one can re-write the matching condition as
\begin{equation}
E_p \epsilon_\gamma \simeq 0.32~ \Gamma {\cal D} ~{\rm GeV}^2~.
\label{resonant1} 
\end{equation} 
Here $\epsilon_\gamma = {\cal D} \epsilon'_\gamma/(1+z)$ is the
observed photon energy, while
$E_p=\Gamma E'_p/(1+z)$ is the energy of the proton as measured by
an earth observer, if it could escape the source (instead of producing
$\pi_0$ photons) and reach earth without energy loss. This is because
the proton energy in 
the rest frame of the AGN central engine (in which the jet is observed 
to move with a Lorentz factor $\Gamma$) is $\Gamma E'_p$. An earth 
observer is at rest of this frame, but with an additional effect due
to cosmic expansion. This definition of $E_p$ is not of significance
in calculating the TeV photon spectrum, but is more convenient to
discuss UHECRs (see III). 
In (\ref{resonant1}), the $(1+z)$ parameter has been neglected 
due to the small redshift $z = 0.002$ of the source.

In the comoving frame, each pion carries $\sim 0.2$ of the proton energy.
Considering each $\pi^0$ splits into two $\gamma$-rays, the $\pi^0$-decay
$\gamma$-ray energy in the observer frame can be written as
$E_\gamma = {\cal D} E'_p/10 = ({\cal D}/10\Gamma) E_p$. The
matching condition between the $\pi^0$-decay photon energy and the 
target photon energy is therefore
\begin{equation}
E_\gamma \epsilon_\gamma \simeq 0.032~ {\cal D}^2 ~{\rm GeV}^2~,
\label{resonant2} 
\end{equation} 

Modeling Cen A 
suggests a viewing angle $\theta_{ob}$ between
$15^{\rm o}$ to $80^{\rm o}$ \cite{Tingay:1998aa,Roustazadeh:2011zz}.
The Doppler factor ${\cal D}$ is found in the range of 
0.12 - 3.7  \cite{Chiaberge:2001ek,Abdo:2010fk,Roustazadeh:2011zz,arXiv:1107.5576} . 
In the following, we adopt a nominal model of \cite{Abdo:2010fk},
with the following parameters: ${\cal D}=1$, $\Gamma=7$, the
comoving blob size ${\cal R}'_b=3\times 10^{15}$ cm, and the comoving
magnetic field strength $B'=6.2$ G, but keep parameter dependences in
the formulae.
This model fits well (the green curve in Fig.\ref{SED}) the bulk of 
observed photon flux starting from the low energy regime to 10s of GeV.
The second SED peak (SSC) is at $\epsilon^p_\gamma \simeq 
170$ keV with the corresponding observed photon energy flux
$F_{\gamma}(\epsilon^p_\gamma)
=9.0\times 10^{-10}~{\rm erg~cm^{-2}~s^{-1}}$ \cite{Steinle:1998}.
It is clearly shown from Fig.\ref{SED} that the model cannot account for
the observations in higher energies.

Adopting ${\cal D}=1$, the target peak photon energy $\epsilon^p_\gamma
=170$ keV is matched by $E^p_p \sim 13$ TeV and $E^p_\gamma 
\sim 190$ GeV for the $\Delta$-resonance condition (Eq.(\ref{resonant1})
and (\ref{resonant2})). The typcial photon energy $E^p_\gamma$ is in 
the energy range of HESS detection.
The optical depth of the $\Delta$-resonance
process in the emission region can be estimated as
$\tau_{p\gamma}=n'_{\gamma}\sigma_{\Delta} {\cal R}'_b$, where
$n'_{\gamma}$ is the comoving photon number density in the jet, which 
is given as $n'_{\gamma} = \eta (L_{\gamma}/{\cal D}^\kappa)/[4\pi
{\cal R'}^2_b c (\epsilon_\gamma/{\cal D})]$, with $\kappa \sim (3-4)$
(depending on whether the jet is continuous or discrete) and $\eta \sim 1$. 
At $\epsilon^p_\gamma=170$ keV, the observed photon luminosity is 
$L_{\gamma}(\epsilon^p_\gamma)
=1.32\times 10^{42}~ (d/3.5~{\rm Mpc})^2 ~{\rm erg~s^{-1}}$. 
For ${\cal D}=1$, this gives 
$n'_{\gamma}(\epsilon^p_\gamma) \sim 1.4 \times 10^6 ~{\rm cm}^{-3}
(d/3.5~{\rm Mpc})^2 ({\cal R'}_{b,15.5})^{-2}{\cal D}^{-\kappa+1}$ and 
$\tau_{p\gamma}(\epsilon^p_\gamma) \sim 2.1 \times 10^{-6}
(d/3.5~{\rm Mpc})^2 ({\cal R'}_{b,15.5})^{-1}{\cal D}^{-\kappa+1}$.

The photon energy flux $F_\gamma(E^p_\gamma)$ (effectively the 
$E^2 dN/dE$ spectrum at $E=E^p_\gamma$) is related to the total proton 
number in the source. The total electron number can be constrained
based on the synchrotron+SSC modeling. However, since we do not know
the composition of the jet, especially the lepton-to-proton number 
ratio (pair multiplicity), one cannot calculate $F_\gamma(E^p_\gamma)$
from the available data. We therefore derive it through fitting the
high energy photon spectrum.

Once $F_\gamma(E^p_\gamma)$ is fitted from the data, one can calculate
the spectrum of the $\pi^0$-decay hadronic component, which depends
on the spectra of the protons and of the target photons.
We assume that protons have a power law distribution
$N(E_p) dE_p \propto E_p^{-\alpha} dE_p$. Since the number of $\pi^0$-decay
photons at a particular energy depends on the number of protons and optical
depth, i.e. $N(E_\gamma) \propto N(E_p) /\tau_{p\gamma} \propto
N(E_p) n'_\gamma(\epsilon_\gamma)$ (where $E_\gamma$, $E_p$, $\epsilon_\gamma$
satisfy the matching conditions (\ref{resonant1}) and (\ref{resonant2})), one
can calculate the $\pi^0$-decay spectrum through the scaling
\be
\frac{F_\gamma(E_\gamma)}{F_\gamma(E^p_\gamma)} = \frac{n'_\gamma(\epsilon_\gamma)}
{n'_\gamma(\epsilon^p_\gamma)} \left(\frac{E_\gamma}{E^p_\gamma}\right)^{-\alpha+2}~,
\label{spectrum}
\ee
where we have used the relation $E_p/E^p_p=E_\gamma/E^p_\gamma$, and the power 
index $+2$ converts the photon number spectrum to energy spectrum. We fix the
proton spectral index to $\alpha = 3.08$.
This index is found to well fit the HESS data in the high energy
regime, and it is 
also the typical cosmic ray spectral index in the UHECR regime. At the
energy $E_p^p\sim$ 13 TeV (which corresponds to the $E_\gamma^p \sim$ 190 GeV
peak of the hadronic component), the proton luminosity at $\sim 13$ TeV 
is 
$L_p(E_p^p) \sim (15/2) L_\gamma(E_\gamma^p) [\tau_{p\gamma}
(\epsilon_\gamma^p)]^{-1} \sim 4.0 \times 10^{45}~{\rm erg~s^{-1}}
({d}/{3.5~{\rm Mpc}})^{-2} {\cal R'}_{b,15.5}{\cal D}^{\kappa-1}$.
This is smaller than the Eddington luminosity of the central black hole
$L_{\rm Edd} \sim 1.3 \times 10^{46}~{\rm erg~s^{-1}} (M/10^8 M_\odot)$.
In order not to violate the energy budget constraint posted by the 
Eddington luminosity, it is required 
that the proton energy spectrum should break to a harder index (e.g. 
$\alpha \sim 2$) at low energies. In our calculation, we introduce 
a break energy $E_p^b$, so that $\alpha=2$ for 
$E_p < E_p^b$, and $\alpha=3.08$ for $E_p > E_p^b$. 
We numerically calculate the model spectrum. 
As can be seen from Fig.\ref{SED}, 
the hadronic model spectrum, along with the leptonic spectrum, can
well interpret the observational data above GeV for a wide range of
$E_p^b$ values (from 4-25 TeV).

For $E_{\gamma} <  E^p_{\gamma}$, the $\Delta$-resonance matching condition
(Eq.(\ref{resonant2})) requires $\epsilon_\gamma > \epsilon^p_\gamma = 170$ 
keV. The drop of the target photon flux then results in a decreasing photon
flux. 
The harder proton spectrum below $E_p^b$ strengthens the effect.

The same applies to $E_\gamma > E^p_\gamma$, whose target photons
have $\epsilon_\gamma < \epsilon^p_\gamma$. Since the number of protons
increases with decreasing energy (power law distribution), the real energy
flux peak of the hadronic component is slightly smaller than $E^p_\gamma
=190$ GeV.

{\em III. UHECR flux.} The same model can be used to estimate the
expected UHECR flux from Cen A. The maximim
energy to which cosmic rays can be accelerated is constrained by the
size of the emitting region and the magnetic field in it. For Cen
A, one has \cite{Dermer:2008cy}
\be
E_{p,max}=4\times 10^{19} \left ( \frac{ B'}{6.2\, G}  \right )
\left ( \frac{ t_v}{10^5 s}  \right ) \left ( \frac{ \Gamma}{7}
\right ) ~{\rm eV},
\ee
where $t_v\sim1$ day is the observed variability timescale, which 
determines the size of the emission region, and the best-fit values of
$B'$ and $\Gamma$ of the leptonic model parameters have been adopted. 
Above this energy the number of cosmic rays should follow an exponential 
decay. Bearing in mind the uncertainties in the viewing angle, it is
possible that the maximum proton energy can reach 57 EeV or even higher
for a same Doppler factor ${\cal D}$ (but with a larger $\Gamma$). In
the following, we assume that $E_{p,max}$ is extended to 57 EeV.

Within our model, based on the flux at $E^p_\gamma$ one can estimate 
the cosmic ray flux at $E^p_p$.
One out of $\tau^{-1}_{p\gamma}$ protons interact with the target photons to
produce gamma-rays through $\pi^0$ decay. So the 
proton flux $F_{p}(E_p)$ at proton energy $E_p$ is related to the 
high energy photon flux $F_{\gamma}(E_\gamma)$ at the photon energy 
$E_\gamma$ through
\be
F_p(E_p) = 7.5 F_\gamma(E_\gamma) [\tau_{p\gamma}(E_p)]^{-1}~.
\ee
The factor 7.5 comes from the fact that the $\Delta$-resonance has 2/3 
probability to decay to the $p\pi^0$ channel as shown in Eq.(\ref{decaymode}), 
and each pion carries 20\% of the proton
energy. At $E^p_{\gamma}= 190$ GeV, the best fit model flux is $F_{\gamma}
(E^p_\gamma)\sim 6.6\times 10^{-13}~{\rm erg~s^{-1}~cm^{-2}}$.
Since the proton energy flux $F_p \propto E^{-\alpha+2}_p$, we obtain the
source proton luminosity at any energy $E_{p}$ through
\be
\frac{F_p(E_{p})}{F_p(E^p_{p}) }=
\left (\frac{E_{p}}{E^p_{p}}\right )^{-\alpha+2}~.
\label{plum}
\ee
Plugging in $E_p = 57$ EeV, and $E^p_p = 13$ TeV, we obtain the UHECR
flux above 57 EeV as $F_p(57~{\rm EeV}) = 1.6\times 10^{-13}
~{\rm erg~s^{-1}~cm^{-2}}$.

The Pierre Auger Observatory (PAO) reported 
that there are roughly 10 UHECR events above 57 EeV 
concentrated around the Centaurus direction, a region with a high AGN
density \cite{Abraham:2007si, Abraham:2010yv, Lemoine:2009zza}.
Two of these events were found to fall within 3 degrees from Cen A
\cite{Auger:2009}, suggesting the evidence
that Cen A may be the first UHECR source.   

We can estimate the expected number of UHECR proton events above 57 EeV 
detectable by the PAO array \cite{Abraham:2007si, Abraham:2010yv}. 
Taking Cen A as a point source, the integrated exposure of PAO  is
$\Xi= 9000/\pi~ {\rm km}^2$. One has to also consider the relative exposure 
$\omega(\delta)$ for the angle of declination $\delta$. For Cen A, 
$\delta = 47^{\circ}$, and the corresponding value is $\omega
(\delta)\simeq 0.64$\cite{:2007qd}.  The time duration for data
collection by PAO is about 15/4 yr between 1st January 2004
and August 2007. So the expected total number of UHECR proton event 
above 57 EeV is
\be
\# {\rm  Events}=\frac{\zeta F_p (57 {\text EeV})}{57 \text {EeV}} \Xi  
\omega (\delta) \, \frac{15}{4}{\text yr} =3.7\zeta,
\ee
where $\zeta$ denotes the fraction of UHECRs that can escape from the
source region. We can see for a reasonable value $\zeta \sim 50\%$, the 
predicted value 1.9 matches nicely the detected 2 events from PAO. 
 Interperting the 2 UHECRs as associated to Cen A implies that the
intergalactic magnetic fields have a strength weaker than $10^{-12}$ G.
This is consistent with a dipole extrapolation of the galactic magnetic
fields, but is inconsistent with a wind ($r^{-2}$-dependence)
extrapolation.

{\em VI. High energy neutrinos.}
From the decay mode of Eq.(\ref{decaymode}), we can see that the fluxes 
of neutral and charged pions have the relation $F(\pi^0)=2F(\pi^+)$. 
Since each neutrino shares 1/4 of the $\pi^+$ energy, while each
photon shares 1/2 of the $\pi^0$ energy, the energy relationship 
between a $\gamma$-ray and a neutrino produced by protons of same
energy satisfies $E_{\nu}=E_{\gamma}/2$.
The neutrino energy flux can be estimated as
$F_\nu = (3/4) F_{\pi^+} \sim (3/8) F_{\pi^0} = (3/8) F_\gamma$.
The maximum neutrino flux is therefore 
$F^p_\nu \simeq 2.5 \times 10^{-13}~{\rm erg~s^{-1}~cm^{-2}}
\simeq 1.5 \times 10^{-10} ~{\rm GeV~s^{-1}~cm^{-2}}$,
with a typical energy $E^p_\nu = E^p_\gamma/2 \simeq 95$ GeV.
This flux level is well below the current neutrino flux upper limit
imposed by IceCube \cite{Karg:2010yh}.

{\em VII. Discussion.}
 
We have proposed a hadronic model to interpret both the TeV
data and the two UHECR events detected from Cen A. The model requires
a relatively high proton-to-electron luminosity ratio (of order 
$10^3-10^4$), with the proton luminosity close to the Eddington
luminosity of the black hole. On the other hand, in view of the
sharp dependence of $L_p$ on ${\cal D}$ and the uncertainties 
in modeling of the leptonic component, such a scenario can be
validated for a choice of reasonable parameters.
An alterative way of accounting for the TeV component through a
hadronic component may be through invoking photon-pair cascade 
initiated from UHECR $p\gamma$ interactions \cite{ arXiv:1111.0936}
which is beyond the scope of this Letter.

We thank Markus B\"ottcher and Parisa Roustazadeh for sharing with us
the SED data and their leptonic model curve for Cen A,  Kohta
Murase and Andres Osorio Oliveros for helpful discussion.
This work is partially supported by DGAPA-UNAM (Mexico) Project 
No. IN101409  (SS) and Conacyt Project No. 105033 (NF) and by 
NASA NNX09AO94G and NSF AST-0908362 (BZ).

\end{document}